\documentclass[letterpaper]{sig-alternate}
\usepackage{graphicx,color,colortbl,psfrag,subfigure}
\usepackage[usenames,dvipsnames]{xcolor}
\usepackage{bbding}
\usepackage{subfigure}
\usepackage{tabularx}
\usepackage{multirow}
\usepackage{array}
\usepackage{upgreek}
\usepackage{bigints}

\usepackage{epsfig}
\usepackage{mdwlist}
\usepackage{flushend}
\usepackage{array}
\usepackage{bm}
\usepackage{algorithm}
\usepackage{algpseudocode}

\usepackage{amssymb}
\usepackage{amsmath}
\usepackage{amsfonts}
\usepackage{mathrsfs}

\usepackage[bookmarks=true,
			colorlinks=true,
			linkcolor=black, 
			urlcolor=black,  
			citecolor=black]{hyperref}
\usepackage{url}

\DeclareMathOperator*{\argmin}{arg\,min}
\DeclareMathOperator*{\argmax}{arg\,max}

\usepackage{balance}

\begin{document}
\conferenceinfo{ADKDD'14,}{August 24 - 27 2014, New York, NY, USA}
\CopyrightYear{2014}
\crdata{978-1-4503-2999-6/14/08}


\title{A Dynamic Pricing Model for Unifying Programmatic Guarantee and Real-Time Bidding in Display Advertising} 
\numberofauthors{1} \author{ \alignauthor
  Bowei Chen, Shuai Yuan and Jun Wang\\
  University College London\\
  \email{\{bowei.chen, s.yuan, j.wang\}cs.ucl.ac.uk} } \date{} \maketitle

\begin{abstract}

There are two major ways of selling impressions in display advertising. They are either sold in spot through auction mechanisms or in advance via guaranteed contracts. The former has achieved a significant automation via real-time bidding (RTB); however, the latter is still mainly done over the counter through direct sales. This paper proposes a mathematical model that allocates and prices the future impressions between real-time auctions and guaranteed contracts. Under conventional economic assumptions, our model shows that the two ways can be seamless combined programmatically and the publisher's revenue can be maximized via price discrimination and optimal allocation.  We consider advertisers are risk-averse, and they would be willing to purchase guaranteed impressions if the total costs are less than their private values. We also consider that an advertiser's purchase behavior can be affected by both the guaranteed price and the time interval between the purchase time and the impression delivery date. Our solution suggests an optimal percentage of future impressions to sell in advance and provides an explicit formula to calculate at what prices to sell. We find that the optimal guaranteed prices are dynamic and are non-decreasing over time. We evaluate our method with RTB datasets and find that the model adopts different strategies in allocation and pricing according to the level of competition. From the experiments we find that, in a less competitive market, lower prices of the guaranteed contracts will encourage the purchase in advance and the revenue gain is mainly contributed by the increased competition in future RTB. In a highly competitive market, advertisers are more willing to purchase the guaranteed contracts and thus higher prices are expected. The revenue gain is largely contributed by the guaranteed selling.  
  
\end{abstract}


\section{Introduction}\label{dp:introduction}

Over the last few years, the demand for automation, integration and optimization has been the key driver for making online advertising one of the fastest advancing industries. In display advertising, a significant development is the emergence of real-time bidding (RTB), which allows buying and selling display impressions in real-time and even a single impression at a time~\cite{Google_2011,Yuan_2013_2}. Yet, despite the strong growth of RTB, according to~\cite{eMarketer_2013_RTB}, 75\% of publishers' revenue in 2012 still came from 20\% guaranteed inventories, which were mainly sold through direct sales by negotiation.

Guaranteed inventories stand for guaranteed contracts sold by top tier websites. Generally, they are: highly viewable because of good position and size; rich in the first-party data (publishers' user interest database) for behavior targeting; flexible in format, size, device, etc.; audited content for brand safety. Therefore, it is not surprising that guaranteed inventories are normally sold in bulk at high prices in advance than those sold on the spot market. 

\begin{figure}[t]
\scriptsize
\centering
\includegraphics[width=0.45\linewidth]{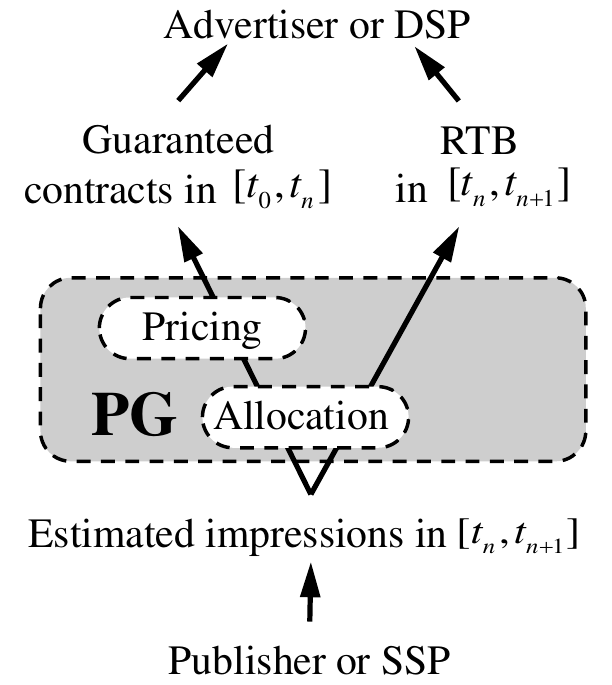}
\vspace{-10pt}
\caption{A systematic view of programmatic guarantee (PG) in display advertising: $[t_0, t_n]$ is the time period to sell the guaranteed impressions that will be created in future period $[t_n, t_{n+1}]$.}
\label{fig:shift_of_selling}
\end{figure}

Programmatic guarantee (PG), sometimes called programmatic reserve/premium~\cite{Dunaway_2012,OpenX_2013}, is a new concept that has gained much attention recently. Notable examples of some early services on the market are \texttt{iSOCKET.com}, \texttt{BuySellAds.com} and \texttt{ShinyAds.com}. It is essentially an allocation and pricing engine for publishers or supply-side platforms (SSPs) that brings the automation into the selling of guaranteed inventories apart from RTB. Figure~\ref{fig:shift_of_selling} illustrates how PG works for a publisher (or SSP) in display advertising. For a specific ad slot (or user tag\footnote{Group of ad slots which target specific types of users.}), the estimated total impressions in a future period can be evaluated and allocated algorithmically at the present time between the guaranteed market and the spot market. Impressions in the former are sold in advance via guaranteed contracts until the delivery date while in the latter are auctioned off in RTB. Unlike the traditional way of selling guaranteed contracts, there is no negotiation process between publisher and advertiser. The guaranteed price (i.e., the fixed per impression price) will be listed in ad exchanges dynamically like the posted stock price in financial exchanges. Advertisers or demand-side platforms (DSPs) can see a guaranteed price at a time, monitor the price changes over time and purchase the needed impressions directly at the corresponding guaranteed prices a few days, weeks or months earlier before the delivery date. 


Developing a revenue maximization model for the programmatic guarantee is sophisticated and challenging. We need to solve the problem of selling unstorable impressions in advance. Similar problems have been studied in many other industries. Examples include retailers selling fashion and seasonal goods and airline companies selling flight tickets~\cite{Talluri_2004}. However, in display advertising, impressions are with uncertain salvage values because they can be auctioned off in real-time on the delivery date. The combination with RTB makes our work interesting and novel. 

Several economic assumptions are made in our study. We consider that future supply and demand of impressions from an ad slot (or user tag) can be well estimated and assume that advertisers' purchase behavior of guaranteed contracts are determined by both the guaranteed price and the time interval between the purchase time and the impression delivery date.
For RTB, we consider the seal-bid second price auction and discuss both probabilistic and empirical distributions of advertisers' bids. 
Under the above assumptions, we develop an algorithmic framework that gives out a functional form of the dynamic optimal price and computes the optimal amount of future impressions to sell in advance. 


We evaluate our development with two RTB datasets. Advertisers bidding behaviors in RTB are investigated and we find that the developed model adopts different strategies in pricing and allocating impressions according to the level of competition on the spot market. If the spot market in future is less competitive, a small amount of impressions would be sold via guaranteed contracts at low prices. The maximized revenue is mainly contributed by the spot market because there is a significant growth in the expected price of auctions in the future. In a highly competitive market, the model allocates more future impressions into guaranteed contracts at high prices and the maximized revenue mainly comes from the guaranteed selling. Under either situation, the revenue can be maximized successfully.

The rest of the paper is organized as follows. Section~\ref{dp:related_work} reviews the related work. In Section~\ref{dp:the_model}, we formulate the problem, discuss our assumptions and provide a solution. Section~\ref{dp:experiments} presents the results of our experimental evaluation  and Section~\ref{dp:conclusion} concludes the paper. 

\begin{table}[t]
\scriptsize
\begin{center}
\caption{Summary of key notations and terminology.}
\label{tab:notations}
\begin{tabular}{p{0.47in}p{2.6in}}
\hline
$t_0, \cdots, t_{n+1}$  & The discrete time points: $[t_0, t_n]$ is the period to sell the guaranteed impressions; $[t_n, t_{n+1}]$ is the period that the estimated impressions should be created, auctioned off (in RTB) and delivered.\\
$t \in [0, T]$ & The continuous time where $t_0 = 0$, $t_n = T$. \\
$\uptau$  & The remaining time to the impression delivery period and $\uptau = T-t \in [0, T]$.\\
$Q$ & The estimated number of total demanded impressions for the ad slot in $[t_n, t_{n+1}]$.\\
$S$ & The estimated number of total supplied impressions for the ad slot in $[t_n, t_{n+1}]$.\\
$p(\uptau)$ & The guaranteed price to sell an impression when the remaining time till the delivery period is $\uptau$.\\
$\theta(\uptau, p(\uptau))$ & The proportion of those who want to buy an impression in advance at $\uptau$ and at $p(\uptau)$.\\
$f(\uptau)$ & The density function so that the number of those who want to buy in advance in $[\uptau, \uptau + d \uptau]$ is $f(\uptau) d\uptau$.\\
$\omega$ & The probability that the publisher fails to deliver a guaranteed impression in the delivery period.\\
$\kappa$ & The size of penalty: if the publisher fails to deliver a guaranteed impression that is sold at $p(\uptau)$, he needs to pay $\kappa p(\uptau)$ penalty to the advertiser.\\
$\xi$ & The number of advertisers who need an impression in RTB, also called the per impression demand.\\
$\phi(\xi)$ & The expected payment price of an impression in RTB for the given demand level $\xi$.\\
$\psi(\xi)$ & The expected risk of an impression in RTB for the given demand level $\xi$.\\
$\lambda$ & The level of risk aversion for advertisers.\\
$\pi(\xi)$ & The expected winning bid of an impression in RTB for the given demand level $\xi$.\\
\hline
\end{tabular}
\vspace{-10pt}
\end{center}
\end{table}

\section{Related Work}\label{dp:related_work}

In online advertising, revenue maximization is always the key issue for publishers, search engines and SSPs. Various attempts have been made to address this challenge, including inventory management and allocation~\cite{Bharadwaj_2010,Roels_2009,Radovanovic_2012}, ads selection and matching~\cite{Broder_2007,Chakrabarti_2008,Yuan_2012}, reserve price optimization~\cite{Ostrovsky_2011,Edelman_2010} and advertising ratio control~\cite{Dewan_2003} etc. In this section we review the related work on guaranteed delivery.

In~\cite{Feldman_2009}, an ads selection and matching algorithm is studied where a publisher\rq{}s objective is not only to fulfill the guaranteed contracts but also to deliver well-targeted impressions to advertisers. The allocation of impressions between guaranteed and non-guaranteed channels is discussed by~\cite{Ghosh_2009}, where a publisher is considered to act as a bidder and to bid for guaranteed contracts. A good property of this setting is that the publisher acts as a bidder would possibly allocate impressions to online auctions only when the winning bids are high enough. The same allocation problem is discussed in~\cite{Balseiro_2011} by using stochastic control models. For a given impression price, the publisher decides whether to send it to ad exchanges or to assign it to an advertiser with a fixed price. The total revenue from ad exchanges and reservations are then maximized. The work of~\cite{Roels_2009} considers a similar framework to~\cite{Balseiro_2011}, where the publisher can dynamically select which guaranteed buy requests to accept and then delivers the guaranteed impressions. Compared to~\cite{Balseiro_2011}, the uncertainty in advertisers\rq{} buy requests and the traffic of website are explicitly modeled in revenue maximization. A lightweight allocation framework is discussed by~\cite{Bharadwaj_2012}. It intends to simplify the computations in optimization and to let real servers to allocate ads efficiently and with little overhead. The work of~\cite{Bharadwaj_2010} proposes two algorithms to calculate the price of selling guaranteed impressions in bulk. However, the effects of online auctions are not considered in their research and the developed algorithms are purely based on the statistics of users' visits to the webpages. 

In the process of providing the guaranteed delivery, a publisher or search engine may also want to cancel the guaranteed contracts if he thinks the non-guaranteed selling is more profitable. In such a situation, the cancellation functionality of a guaranteed delivery system is beneficial to the sell side. Cancellations are discussed in~\cite{Babaioff_2009,Constantin_2009}, where a publisher can cancel a guaranteed contract later if he agrees to pay a penalty. Publishers can enjoy a lot of flexibility with cancellations but there may exist speculators in the game who pursue the penalty only. In fact, the cancellation penalty in online advertising is very similar to the over-selling booking of airline tickets. Several important over-selling booking models are discussed in~\cite{Talluri_2004}. 
 
The so-called ad option contracts are a more flexible guaranteed delivery mechanism in online advertising~\cite{Chen_2013,Moon_2010,Wang_2012_1}. Advertisers are guaranteed a priority buying right (but not an obligation) of their targeted future inventories. They can decide to pay fixed prices to obtain the inventories or rejoin the auctions in the future. The ad option contracts require advertisers to pay an upfront fee first in exchange for the right in the future. In~\cite{Moon_2010,Wang_2012_1}, the ad option contracts allow advertisers to choose a combined payment scheme of fixed prices. For example, an advertiser can pay a fixed cost-per-click (CPC) for display impressions. In~\cite{Chen_2013}, the contracts enable advertisers to target multiple keywords in sponsored search. However, the studies of~\cite{Chen_2013,Moon_2010,Wang_2012_1} focus on the option pricing methods under various objectives and do not discuss how to effectively allocate the inventories.

\section{The Model}\label{dp:the_model}

Consider there is a premium ad slot on a publisher's webpage. If there is a user comes to this webpage, the ad slot can generate a chance of ad view, usually referred to as an impression. In RTB, an impression is auctioned off simultaneously once a user comes and the winning bidder (i.e., the advertiser) has his ad displayed to the user~\cite{Google_2011,Yuan_2013_2}. Suppose that the publisher can estimate supply and demand of impressions from an ad slot (or user tag) from historical transactions and plan to sell some of the future impressions via guaranteed contracts in advance in order to maximize the revenue. We consider an environment that is risk-averse and both publisher and advertiser make their strategies by maximizing their expected utilities~\cite{Bhalgat_2012}. In other words, the advertiser is willing to pay a higher price for a fixed number of future impressions if the delivery is guaranteed. This gives the publishers an additional possibility of increasing their revenue by pre-selling some future impressions, apart from the price discrimination over time.

\subsection{Problem Formulation}
The optimization problem can be expressed as
\begin{align}
\boldsymbol\max & \ \Bigg\{ 
\hspace{-15pt}
\underbrace{\int_{0}^{T} (1- \omega \kappa) p(\uptau) \theta(\uptau, p(\uptau)) f(\uptau) d\uptau}_{\textrm{
\begin{tabular}{l}
$G =$ Expected total revenue from guaranted selling minus\\
expected penalty of failling to delivery\\
\end{tabular} }} \nonumber \\ 
& \ \ \ \ \ \ \ \ \ \ \ \ 
+ \underbrace{\bigg(S - \int_{0}^{T} \theta(\uptau, p(\uptau)) f(\uptau) d\uptau \bigg) \phi(\xi)}_{\textrm{$H =$ Expected total revenue from RTB}}  \ \ \Bigg\}, \label{eq:objective_1} \\[0.1in]
\textbf{s.t.} & \ \ \ 
p(0) = \Bigg\{\hspace{-5pt}
\begin{array}{ll}
\phi(\xi) + \lambda \psi(\xi), & \hspace{-5pt}  \textrm{if } \pi(\xi) \geq \phi(\xi) + \lambda \psi(\xi)\\
\pi(\xi), & \hspace{-5pt}  \textrm{if } \pi(\xi) < \phi(\xi) + \lambda \psi(\xi),\\
\end{array} 
\hspace{-5pt}\label{eq:constraint_1}
\end{align}
where 
\begin{align*}
\xi = & \ \frac{\textrm{Remaining demand in } [t_n, t_{n+1}]}{\textrm{Remaining supply in } [t_n, t_{n+1}]} 
= \frac{Q - \int_{0}^{T} \theta(\uptau, p(\uptau)) f(\uptau) d\uptau }{S - \int_{0}^{T} \theta(\uptau, p(\uptau)) f(\uptau) d\uptau }.
\end{align*}

The notations are given in Table~\ref{tab:notations}. The publisher's expected total revenue contains: the expected revenue from guaranteed impressions sold during $[0, T]$; the expected penalty of failing to delivery guaranteed impressions in $[t_n, t_{n+1}]$; the expected revenue from RTB in $[t_n, t_{n+1}]$; and the price constraint that ensures the advertisers' willingness to buy guaranteed impressions. Eq.~(\ref{eq:constraint_1}) shows that an advertiser's decision of buying either a guaranteed or non-guaranteed impression depends on the expected payment price and his level of risk-aversion.  For simplicity and without loss of generality, we consider each guaranteed impressions as a single guaranteed contract, where in practice this settings can be extended to a bulk sale.

The solution to the above optimization problem appears a bit complicated as it needs to answer how many future impressions to sell and at what prices to sell. Before discussing the solution, we need to make several assumptions, such as the distribution of bids in RTB and the advertisers' purchase behavior in advance.

\begin{algorithm}[!t]
\scriptsize
\caption{\bf Estimate $\phi(\xi)$ by using the robust locally weighted regression (RLWR) method~\cite{Cleveland_1979}.}
\label{algo:rlwr}
\begin{algorithmic}
\Function{\texttt{RLWRSolve}}{$\xi$}
	\State // $(\xi_j, \phi_j)$, $j = 1, \ldots, m$, are the learning data with size $m$.
	\State \hspace{-15pt} $\widehat{\boldsymbol\beta}(\xi) \leftarrow \argmin \sum_{j = 1}^m \varpi_j(\xi)(\phi_j - \beta_0 - \beta_1 \xi_j - \ldots - \beta_d \xi_j^d )^2$,
	\State where $\varpi_j(\xi) \leftarrow \Bigg\{
	\begin{array}{ll}
	\Big(1- \Big|\frac{\xi - \xi_j}{h(\xi)}\Big|^3 \Big)^3, & \textrm{if } \Big|\frac{\xi - \xi_j}{h(\xi)}\Big| < 1, \\
	0, & \textrm{if } \Big|\frac{\xi - \xi_j}{h(\xi)}\Big| \geq 1.\\
	\end{array}
	$
	\State // $h(\xi)$ is the distance from $\xi$ to the most distant neighbor of $\xi$ within the span, and we choose $d=2$.
	\State $\widehat{\phi} \leftarrow \sum_{k = 0}^d \widehat{\beta}_k (\xi) \xi^k$.
	\Loop{$\; i \leftarrow 1$ to $5$ // repeat the update in 5 iterations}
		\State $\boldsymbol{\epsilon} \leftarrow \boldsymbol{\phi} - \widehat{\phi}$, 
		$\chi(\boldsymbol{\epsilon}) \leftarrow \textrm{median}(\mid \boldsymbol{\epsilon} \mid)$.
		\For{$j \leftarrow 1$ to $m$} 
		\State $\varpi_j(\xi) \leftarrow \bigg\{
			\begin{array}{ll}
			\big( 1 - \big(\frac{e_j}{6 \chi(\boldsymbol{\epsilon})} \big)^2 \big)^2, & \textrm{if } \mid \epsilon_j \mid < 6 \chi(\boldsymbol{\epsilon}),  \\
			0, & \textrm{if } \mid \epsilon_j \mid \geq 6 \chi(\boldsymbol{\epsilon}).  \\
			\end{array}
			$
		\EndFor
		\State \hspace{-15pt} $\widehat{\boldsymbol\beta}(\xi) \leftarrow \argmin \sum_{j = 1}^m \varpi_j(\xi)(\phi_j - \beta_0 - \beta_1 \xi_j - \ldots - \beta_d \xi_j^d )^2$.
		\State $\widehat{\phi} \leftarrow \sum_{k = 0}^d \widehat{\beta}_k (\xi) \xi^k$.
	\EndLoop  
	\State \Return $\phi(\xi) \leftarrow \widehat\phi$.
\EndFunction
\end{algorithmic}
\end{algorithm}

\subsection{Distribution of Bids in RTB}

Advertisers bid for individual impressions separately in RTB~\cite{Google_2011,Yuan_2013_2}. Therefore, we consider the following second-price auction: for a single impression from a specific ad slot (or user tag), advertisers submit sealed bids to the publisher (or SSP), and the highest bidder wins the impression but finally pays at the bid next to him. 

We can consider either probabilistic or empirical distribution of bids in RTB. Bidders are assumed to be symmetric in probabilistic method; therefore, advertisers would truthfully bid (i.e, bid at their private values). We adopt the settings in~\cite{Ostrovsky_2011} and assume bids follow the log-normal distribution, denoted by $X \sim \mathbf{LN}(\mu, \sigma^2)$. Then, the expected per impression payment price from a second-price auction is
\begin{align}
\phi(\xi) = & \ \int_{0}^{\infty} x \xi (\xi - 1) g(x)\Big(1 - F(x) \Big) \Big(F(x)\Big)^{\xi-2} dx, \label{eq:exp_price_auction}
\end{align}
where $g(x)$ and $F(x)$ are log-normal density and its cumulative distribution function, respectively, given by
\begin{equation}
g(x) = \frac{1}{x \sigma \sqrt{2 \pi}}\, e^{-\frac{(\ln(x) - \mu)^2}{2\sigma^2}}, 
\ 
F(x) = \frac{1}{2} + \frac{1}{\sqrt{\pi}} \int_0^{\frac{\ln(x) - \mu}{\sqrt{2 \sigma}}} \hspace{-5pt} e^{- z^2} dz,
\nonumber 
\end{equation}
so that $\xi (\xi - 1) g(x) (1 - F(x) ) (F(x))^{\xi-2}$ represents the probability that if an advertiser is the second highest bidder, then one of the $\xi - 1$ other advertisers must bid at least as much as he/she does and all of the $\xi - 2$ other advertisers have to bid no more than he/she does. We can check if bids follow the log-normal distribution by Kolmogorov-Smirnov (K-S)~\cite{Smirnov_1948} and Jarque-Bera (J-B)~\cite{Jarque_1980} statistics (see Table~\ref{tab:bids_distribution_test}). Once the log-normal distribution is met, we can obtain $\phi(\xi)$ numerically because the values of $g(x)$ and $F(x)$ in each integration increment can be calculated. 

If bids do not follow the log-normal distribution, we refer to an empirical method to compute $\phi(\xi)$. Simply, for an ad slot (or user tag), the winning payment prices are trained to develop a regression model that explains their correlation to the level of demand. In this paper, we use the robust locally weighted regression (RLWR) method~\cite{Cleveland_1979} (see Algorithm~\ref{algo:rlwr} and an empirical example in Section~\ref{dp:exp_bid_and_payment_price}). Other statistical learning methods can be developed to estimate $\phi(\xi)$ but we are not going to further investigate them here.

\subsection{Risk Aversion and Purchase Behavior}

Eq.~(\ref{eq:constraint_1}) tells that at time $T$ an advertiser's decision between guaranteed and non-guaranteed channels are indifferent. In this paper, we do not model the advertisers' arrival as a stochastic process~\cite{Talluri_2004}, instead, we consider that the total demand for future impressions is deterministic but can be shift from future to present. The possibility of this shift is because advertisers are assumed to be risk-averse. 

Under our risk aversion settings, $\pi(\xi)$ and $\psi(\xi)$ can be estimated by the RLWR method, and $\lambda$ can be set as any non-negative number. First, the estimation of $\pi(\xi)$ is as same as Algorithm~\ref{algo:rlwr} while we consider highest bids (per transaction) rather than payment prices (per transaction). Second, the estimation of $\psi(\xi)$ is slightly different. We compute a series of standard deviations of daily winning payment prices and use Algorithm~\ref{algo:rlwr} to compute $\psi(\xi)$ for the given demand level. Third, advertisers' risk-averse preference are not same; therefore, $\lambda$ can be regarded as the average risk-aversion level of all advertisers or of key advertisers (we consider the former in the experiments). The larger $\lambda$ the more risk-averse advertisers. More detailed discussion about the estimation of $\pi(\xi)$, $\psi(\xi)$ and $\lambda$ is given in Section~\ref{dp:exp_bid_and_payment_price}. 

Similar to flight tickets booking~\cite{Anjos_2004,Anjos_2005,Malighetti_2009}, we have the following two economic assumptions on demand:
\begin{description}\vspace{-7pt}
\item[\textbf{A-1}] Demand is negatively correlated with guaranteed price as advertisers would buy less impressions if price increases. Given $\uptau$ and $0 \leq p_1 \leq p_2$, then $\theta(\uptau, p_1) \geq \theta(\uptau, p_2)$, subject to the boundary condition $\theta(\uptau, 0) = 1$.\vspace{-7pt}
\item[\textbf{A-2}] Demand is negatively correlated with the time interval between purchase and delivery because more advertisers' would want to buy impressions when the delivery date is approached. Given $p$ and $0 \leq \uptau_2 \leq \uptau_1$, then $\theta(\uptau_2, p) \geq \theta(\uptau_1, p)$. 
\end{description}

We adopt the functional forms of demand proposed by~\cite{Anjos_2004} (which are used in flight tickets booking):
\begin{align}
\theta(\uptau, p(\uptau)) = & \ e^{ - \alpha p(\uptau) (1 + \beta \uptau)}, \label{eq:demand_function_1}\\
f(\uptau) = & \ \zeta e^{- \eta \uptau}, \label{eq:demand_function_2}
\end{align}
where $\alpha$ is the level of price effect, $\beta$ and $\eta$ are the levels of time effect, and the demand density rises to a peak $\zeta$ on the delivery date. Therefore, $f(\uptau) d \uptau$ shows the number who would be willing to purchase in advance, and $\theta(\uptau, p(\uptau))$ represents the proportion of advertisers who want to purchase an impression in advance at $\uptau$ and at $p(\uptau)$. 

\subsection{Optimal Dynamic Prices}

\begin{algorithm}[!t]
\scriptsize
\caption{\bf Solution to Eq.~(\ref{eq:objective_1}).}
\label{algo:solution}
\begin{algorithmic}
\Function{\texttt{PGSolve}}{$\alpha, \beta, \zeta, \eta, \omega, \kappa, \lambda, S, Q, T$}
	\State $\boldsymbol{t} \leftarrow [t_0, \cdots, t_n]$, $0 = t_0 < t_1 < \cdots < t_n = T$.
	\State $\boldsymbol{\uptau} \leftarrow T - \boldsymbol t$, 
	$m \leftarrow 500$.
	\Loop{$\; i \leftarrow 1$ to $m$}		
		\State $\gamma_i \leftarrow \texttt{RandomUniformGenerate}([0,1])$
		\State $\int_{0}^{T} \theta(\uptau, p(\uptau)) f(\uptau) d \uptau \leftarrow \gamma_i S$
		\State $\xi_i \leftarrow (Q - \gamma_i S)/(S -\gamma S)$
		\State $H_i \leftarrow (1-\gamma_i)S \phi(\xi_i)$
		\State $G_i \leftarrow \int_{0}^{T} (1- \omega \kappa) p(\uptau) \theta(\uptau, p(\uptau)) f(\uptau) d\uptau$ 
		\vspace{-5pt} 
		\begin{align}
   		\boldsymbol{p}_i \leftarrow & \boldsymbol\argmax \ G_i, \label{eq:objective_2} \\
                      & \textbf{s.t.} \int_{0}^{T} \theta(\uptau, p(\uptau)) f(\uptau) 					             d\uptau = \gamma_i S, \\
              &  p(0) = \Bigg\{\hspace{-5pt}
\begin{array}{l}
\phi(\xi_i) + \lambda \psi(\xi_i), \textrm{if } \pi(\xi_i) \geq \phi(\xi_i) + \lambda \psi(\xi_i),\\
\pi(\xi_i), \textrm{if } \pi(\xi_i) < \phi(\xi_i) + \lambda \psi(\xi_i).\\
\end{array} 
\hspace{-5pt}
		\end{align}
		\State $R_i \leftarrow \boldsymbol\max G_i + H_i$
	\EndLoop
	\State $\gamma^* \leftarrow \boldsymbol\argmax_{\gamma_i \in \Omega(\gamma)} \{R_1, \ldots,  R_m\}$	
	\State $\boldsymbol{p}^* \leftarrow \boldsymbol\argmax_{\boldsymbol{p}_i \in \Omega(\boldsymbol{p})} \{R_1, \ldots,  R_m\}$	
	\State	\Return $\gamma^*, \mathbf{p^*}$
\EndFunction
\end{algorithmic}
\end{algorithm}

The optimization problem in Eq.~(\ref{eq:objective_1}) can be solved by Algorithm~\ref{algo:solution}. We simulate many values of $\gamma_i \in [0,1], i =1,\ldots, m$. For each given $\gamma_i$, we solve the optimization problem in Eq.~(\ref{eq:objective_2}), find the optimal series of guaranteed prices and calculate the optimal total revenue $R_i$. Then, in the global comparison, we can find the optimal $\gamma^*$ that generates the maximum value of total revenue. 

Let us discuss how to solve the optimization problem in Eq.~(\ref{eq:objective_2}). We consider the following Lagrangian:
\begin{align}
\mathcal{L}(\widetilde\lambda, p(\uptau)) = 
& \ \int_{0}^{T}(1- \omega \kappa) p(\uptau) \theta(\uptau, p(\uptau)) f(\uptau) d\uptau \nonumber \\
& \ \ \ \ + \widetilde\lambda \bigg( \gamma_i S - \int_{0}^{T} \theta(\uptau, p(\uptau)) f(\uptau) d \uptau\bigg), \label{eq:lagranage_function}
\end{align}
where $\widetilde\lambda$ is the Lagrange multiplier. The Euler-Lagrange condition is $\partial \mathcal{L}/\partial p = 0$. For $\uptau \in (0, T]$, we have
\begin{equation}
(1- \omega \kappa) \theta(\uptau, p(\uptau)) + \Big( (1- \omega \kappa) p(\uptau) - \widetilde\lambda \Big) \frac{\partial \theta(\uptau, p(\uptau))}{\partial p(\uptau)} = 0. \label{eq:lagranage_condition}
\end{equation}

Substituting Eq.~(\ref{eq:demand_function_1}) into Eq.~(\ref{eq:lagranage_condition}) then gives the formula of the optimal guaranteed price:
\begin{align}
p(\uptau) = & \ \frac{\widetilde\lambda}{1- \omega \kappa} + \frac{1}{ \alpha (1 + \beta \uptau)}. \label{eq:guaranteed_price}
\end{align}

Consider a small time step $d\widetilde\uptau$, then in $[0, 0+d\widetilde\uptau]$, there are $\theta(0, p(0) f(0) d\widetilde\uptau$ demand fulfilled. Therefore, we have
\begin{align}\label{eq:lagranage_condition_2}
\int_{d\widetilde\uptau}^{T} \theta(\uptau, p(\uptau)) f(\uptau) d\uptau = \gamma_i S - \theta(0, p(0)) f(0) d\widetilde\uptau
\end{align}
By substituting Eqs.~(\ref{eq:demand_function_1})-(\ref{eq:demand_function_2})~\&~(\ref{eq:constraint_1}) into Eq.~(\ref{eq:lagranage_condition_2}), we have
\begin{align}
& \ \frac{-\zeta (1-\omega \kappa )e^{-\left(\frac{\alpha \widetilde{\lambda} \beta }{1-\omega \kappa }+1\right)}}{\alpha \widetilde{\lambda} \beta +(1-\omega \kappa )\eta }\left(e^{-\left(\frac{\alpha \widetilde{\lambda} \beta }{1-\omega \kappa }+\eta \right)T}-e^{-\left(\frac{\alpha \widetilde{\lambda} \beta}{1-\omega \kappa }+\eta\right)d\widetilde{\uptau}}\right) \nonumber \\
= & \ \gamma_iS-e^{-\alpha p(0)} \zeta d\widetilde{\uptau}. \label{eq:lambda_equation}
\end{align}

\begin{figure}[t]
\scriptsize
\centering
\includegraphics[width=0.92\linewidth]{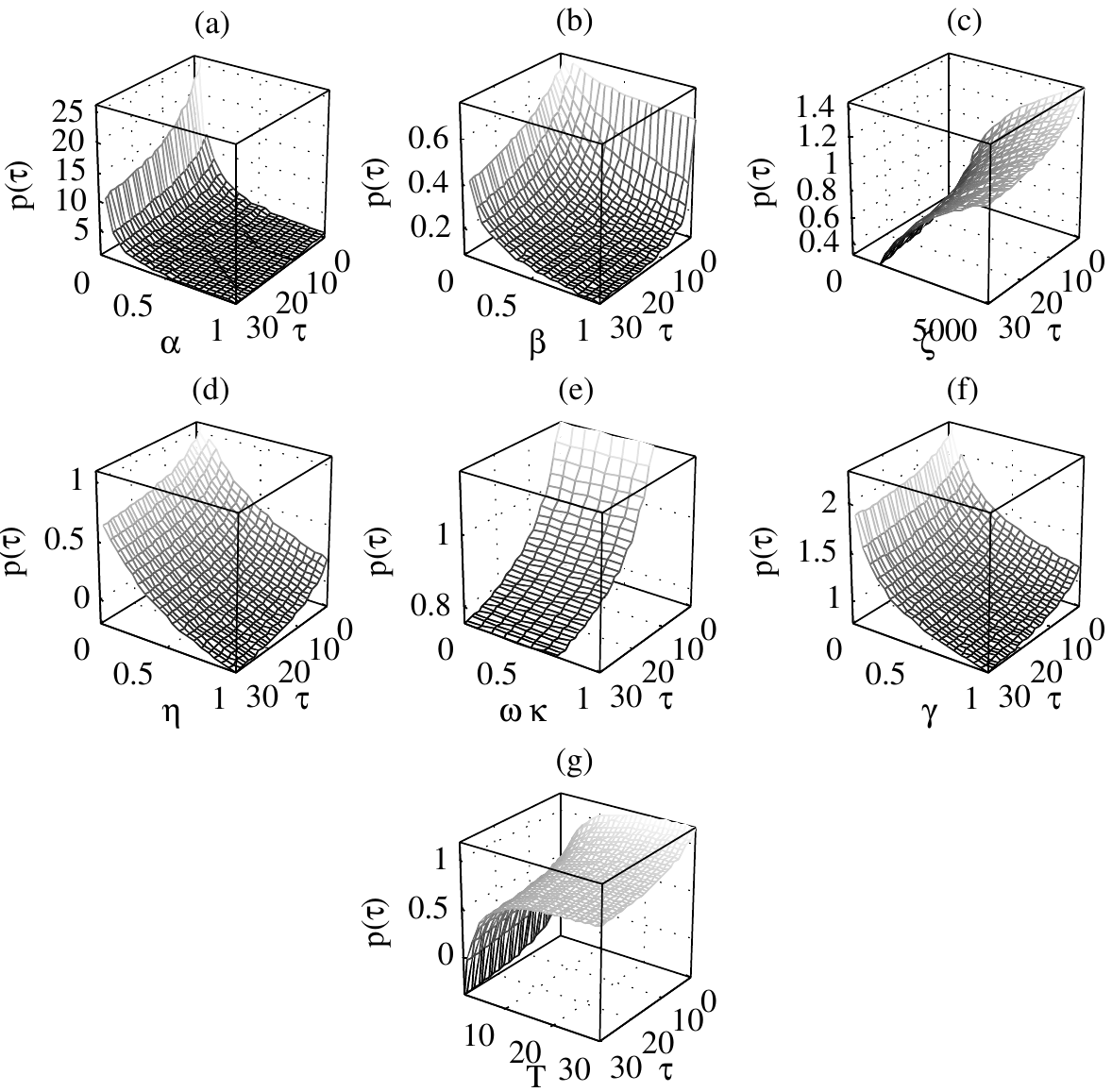}\vspace{-5pt}
\caption{The impact of model parameters on the guaranteed selling prices: $\alpha$, $\beta$, $\zeta$, $\eta$ are defined in Eqs.~(\ref{eq:demand_function_1})~\&~(\ref{eq:demand_function_2}); $\omega \kappa$ is the expected size of penalty; $\gamma$ is the percentage of estimated future impressions to sell in advanced; $T$ is the length of guaranteed selling period; $\uptau$ is the remaining time to the delivery date; $p(\uptau)$ is the guaranteed selling price at $\uptau$.}
\label{fig:parameters_effect}
\begin{minipage}{1\linewidth}
\begin{table}[H]
\centering
\scriptsize
\caption{Summary of RTB datasets.}
\label{tab:datasets}
\begin{tabular}{r|r|r}
\hline
Dataset & SSP & DSP\\
\hline
From    & 08/01/2013 & 19/10/2013\\
 To     & 14/02/2013 & 27/10/2013\\
\# of ad slots  & 31 & 53571  \\
\# of user tags & NA & 69     \\
\# of advertisers   & 374 & 4 \\ 
\# of impressions   & 6646643 & 3158171\\ 
\# of bids      & 33043127   &  11457419 \\
Bid quote       & USD/CPM  & CNY/CPM\\ 
\hline
\end{tabular}
\caption{Experimental design of the SSP dataset.}
\label{tab:ssp_dataset}
\begin{tabular}{r|rr}
\hline
                          & From & To\\
\hline                          
Training set           & 08/01/2013  & 13/02/2013\\
Development set        & 08/01/2013  & 14/02/2013\\
Test set                & \multicolumn{2}{c}{14/02/2013 }\\ 
\hline
\end{tabular}
\vspace{2pt}
\end{table} 
\end{minipage}
\end{figure}

Eq.~(\ref{eq:lambda_equation}) shows that the value of $\widetilde\lambda$ is dependent on $\gamma_j S$ and other parameters; however, the explicit solution of $\widetilde\lambda$ cannot be deduced. We can approximate the value of $\widetilde\lambda$ by using the numerical methods (e.g. the Newton-Raphson method) and rewrite Eq.~(\ref{eq:guaranteed_price}) as follows
\begin{align}
p(\uptau) = & \ \frac{\widetilde\lambda(\alpha, \beta, \zeta, \eta, \omega, \kappa, \gamma_i S)}{1- \omega \kappa} + \frac{1}{ \alpha (1 + \beta \uptau)}. \label{eq:upper_bound_generic}
\end{align}
The notation $\widetilde\lambda(\alpha, \beta, \zeta, \eta, \omega, \kappa, \gamma_i S)$ represents the dependency relationship among $\widetilde\lambda$ and other parameters. Figure~\ref{fig:parameters_effect} gives a numerical investigation on the relationships between $p(\uptau)$ and model parameters. Recall that in Eqs.~(\ref{eq:demand_function_1})-(\ref{eq:demand_function_2}) a large value of $\alpha$ means advertisers are price sensitive; therefore, $p(\uptau)$ decreases if $\alpha$ increases. Similar negative correlations are with $\beta$ and $\eta$. These two parameters describe the time effect on advertisers' willingness to purchase. The model thus encourages advertisers to purchase in advance by selling guaranteed contracts at low prices. Conversely, the optimal price is positively correlated with $\zeta$ because the parameter shows the maximum number of advertisers that would be willing to buy guaranteed impressions at a time point. More advertisers means more competition; therefore, more advertisers would purchase in advance in order to secure the targeted impressions. In such a situation, the model gives out high guaranteed prices and allocates more impressions to guaranteed contracts. While the expected penalty $\omega \kappa$ has less effect on price, the larger  $\omega \kappa$ the higher $p(\uptau)$. It is worth noting that $\omega$ and $\kappa$ are considered as given parameters in this paper because: (i) $\kappa$ can be set by negotiation between publisher and advertiser; (ii) $\omega$ can be estimated\footnote{$\omega$ can be approximated by the percentage of guaranteed impressions that the publisher fails to delivery.} and updated once the PG system runs for a certain period of time. In this paper, we set $\omega = 0.05$, $\kappa = 1$. With less and less supplied impressions to sell on the market, the price $p(\uptau)$ increases. The total length of time period to sell guaranteed contracts positively affects the guaranteed price, the longer $T$, the higher the $p(\boldsymbol\uptau)$.

\section{Experimental Evaluation}\label{dp:experiments}

We describe our datasets in Section~\ref{dp:exp_datasets}, investigate the RTB campaigns in Sections~\ref{dp:exp_bidding_behaviours}-\ref{dp:exp_supply_demand}, discuss the estimation of model parameters in Sections~\ref{dp:exp_bid_and_payment_price}-\ref{dp:exp_demand_guaranteed_imps}, and evaluate the performance of revenue maximization in Section~\ref{dp:exp_reve_analysis}.

\subsection{Datasets}\label{dp:exp_datasets}

We use two different RTB datasets: one from a medium-sized SSP in the UK and the other from a DSP in China. Table~\ref{tab:datasets} shows a brief summary of these two datasets. The SSP dataset is used throughout the whole experiments while the DSP dataset is used for further exploring advertisers' strategies in RTB. In these two datasets, all the bids are expressed as cost per mille (CPM), i.e., the measurement corresponds to the value of 1000 impressions. 

Table~\ref{tab:ssp_dataset} illustrates our experimental design, where the SSP dataset is divided into one training set, one development set and one test set. In the training set, we investigate RTB campaigns and estimate model parameters. In the development set, we use the discussed model to allocate and price the impressions that are created on 14/02/2013. Guaranteed contracts are sold over the period from 08/01/2013 to 13/02/2013 and the rest impressions are auctioned off on the delivery date 14/02/2013. In the development set, we simulate the transactions of guaranteed contracts and calculate the remaining campaigns of RTB on 14/02/2013. The test set contains the actual bids and winning payment prices of 14/02/2013, which is used to evaluate the revenue maximization performance. Note that time periods of training and development sets can be different. For example, the development period can be a few days/weeks later than the training period. However, this requires a number of forecasting methods to estimate all the model parameters (features). As our primary intention here is not to discuss better forecasting methods, we choose a learning period that is close to the impression delivery date so that the learned parameters are more accurate for the evaluation purpose.

\subsection{Bidding Behaviors}\label{dp:exp_bidding_behaviours}

\begin{figure}[t]
\scriptsize
\centering
\includegraphics[width=0.95\linewidth]{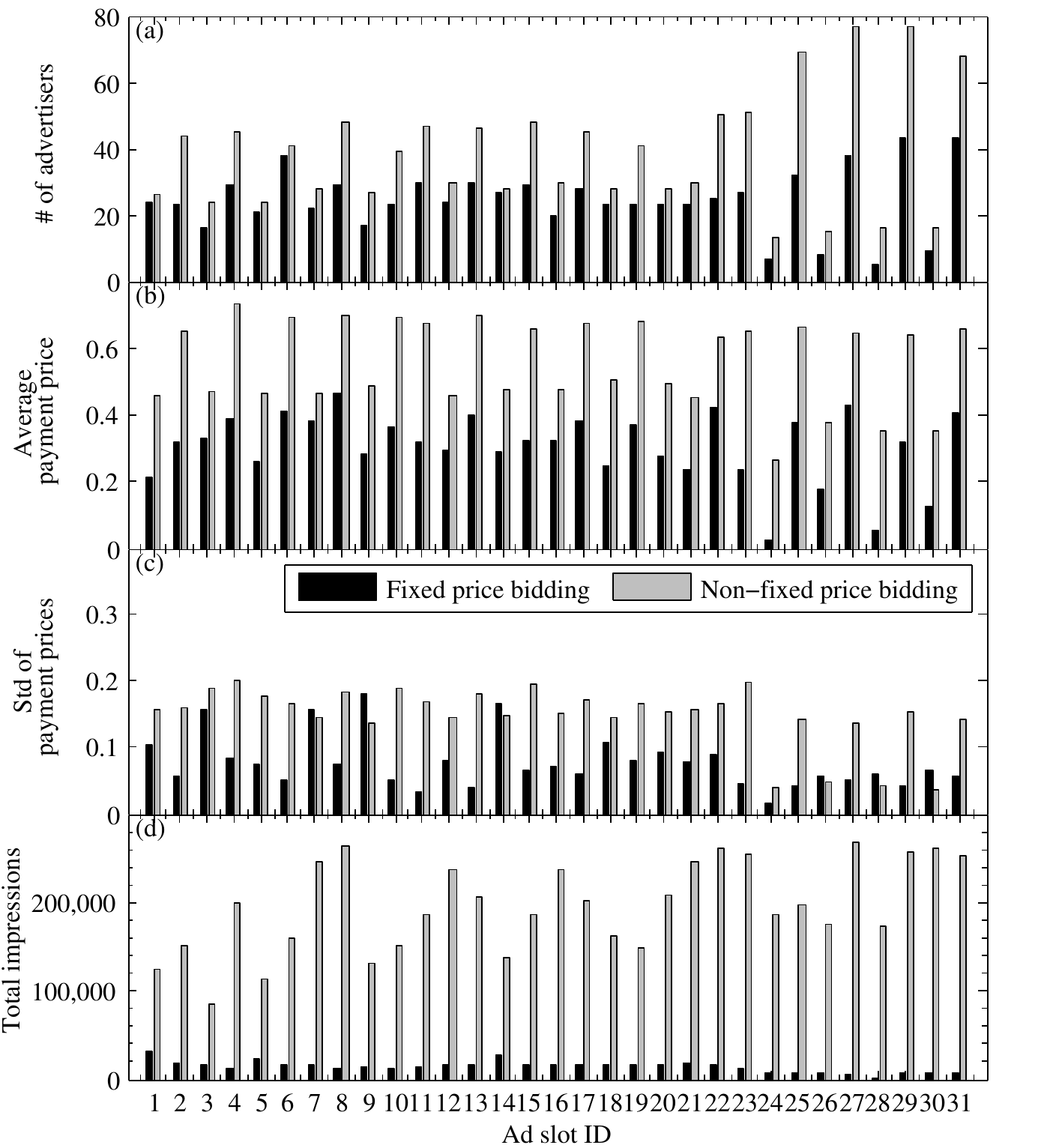}
\vspace{-15pt}
\caption{\hspace{-7pt} Overview of the winning advertisers' statistics from the SSP dataset in the training period.}
\label{fig:advertiser_stats}
\end{figure}

\begin{table}[t]
\scriptsize
\centering
\caption{Summary of the winning advertisers' statistics from the SSP dataset in the training period: the numbers in the brackets represent how many advertisers who use the combined bidding strategies.}
\label{tab:bids_ssp}
\begin{tabular}{r|rrrr}
\hline
Bidding   & \hspace{-10pt} \# of        &  \# of         & \hspace{-10pt} Average & Ratio of \\
strategy  & advertisers  & imps          & change   & payment \\
	      &              & won           & rate of  & price to \\
	      &              &               & payment  & winning \\
	      &		         &		         & prices	& bid\\ 
\hline
Fixed price &  188 (51) &  454681  & 188.85\% & 43.93\% \\
\hspace{-15pt} Non-fixed price &  200 (51) & \hspace{-5pt} 6068908  & \hspace{-5pt}  517.54\% & 58.94\% \\
\hline
\end{tabular}
\vspace{2pt}
\caption{Summary of advertisers' winning campaigns from the DSP dataset. All the advertisers use the fixed price bidding strategy. Each user tag contains many ad slots and an ad slot is sampled from the dataset only if the advertiser wins more than 1000 impressions from it.}
\label{tab:bids_dsp}
\begin{tabular}{r|rrrrr}
\hline
\hspace{-10pt} Advertiser & \# of     & \# of     & \# of   & Average & Ratio of\\
ID      & user      & ad slots  & imps    & change & payment \\
	    & tags	    &           & won     & rate of & price to \\
	    &           &           &         & payment   & winning \\
	    &		    &		    &		  &	prices	    & bid\\ 
\hline
1 & 69 & 635 &  196831 & 58.57\% & 36.07\%\\
2 & 69 & 428 & 144272  & 58.94\% & 34.68\%\\
3 & 69 & 1267& 123361  & 79.24\% & 30.89\%\\
4 & 65 & 15  & 3139    & 104.19\% & 22.32\%\\
\hline
\end{tabular}
\vspace{5pt}
\end{table}

We first examine if selling guaranteed impressions in advance can be a viable way to segment advertisers according to their bids, and then discuss how much of revenue growth can be expected.

Let us first look at advertisers behaviors in RTB. From the SSP dataset, we find that advertisers mainly join auctions in the morning from 6am to 10am. It is the time period that supplied impressions arrive peak. We also find that the winning advertisers' final payments are much less than their bids. Figure~\ref{fig:advertiser_stats} gives some descriptive statistics about this finding across all 31 ad slots. We simply divide the winning advertisers into two groups. The first group contains those who always offer a fixed bid; the second group contains those who frequently change their bids. Figure~\ref{fig:advertiser_stats} shows that more winning advertisers adopt the non-fixed price bidding in RTB. They intend to offer higher bids on each impression, endure more variance in payment prices due to the second price auction, and obtain more impressions. The second price auction in RTB provides an opportunity for making more revenue by selling impressions in advance: 1) a risk-averse advertiser is willing to buy in advance to lock in the price; 2) the publisher would be able to increase the price for the guaranteed contracts by charging advertisers their private valuations rather than the second price bids. The question is how big the difference between the top bids and actually payments (the second price).  Table~\ref{tab:bids_ssp} shows that the publisher can expect 100\% increase in revenue because the current average ratio of actual payment price (the second price) to winning bid (the first price) is about 50\%.

We further examine the DSP dataset, and find all 4 advertisers use the fixed price strategy in their bidding. This might be because the DSP itself adopts the fixed price strategy for these 4 advertisers. While the DSP dataset itself is biased, we can still take a look at the average volatility of the advertisers\rq{}s payment prices and the average ratio of payment price to winning bid. In the DSP dataset, we see an advertiser actually bids for an user tag instead of a specific ad slot. Each user tag contains a set of ad slots that have similar features and can allow the advertiser to target a certain group of online users. In short, RTB is the user targeting bidding. Consider in an user tag the advertiser\rq{}s bids are not well distributed among ad slots, we only investigate the ad slots where the advertiser wins more than 1000 impressions. Table~\ref{tab:bids_dsp} confirms our earlier statement from a buy side perspective. Even using the fixed price bidding strategy, advertisers\rq{} payment prices are volatile (more than 50\% from each impression). In fact, these advertisers can afford more to reduce the risk because the current payment prices are much lower than their private valuations (around 30\% across 4 advertisers).

\subsection{Supply and Demand}\label{dp:exp_supply_demand}

\begin{figure}[t]
\scriptsize
\centering
\vspace{-7pt}
\includegraphics[width=0.95\linewidth]{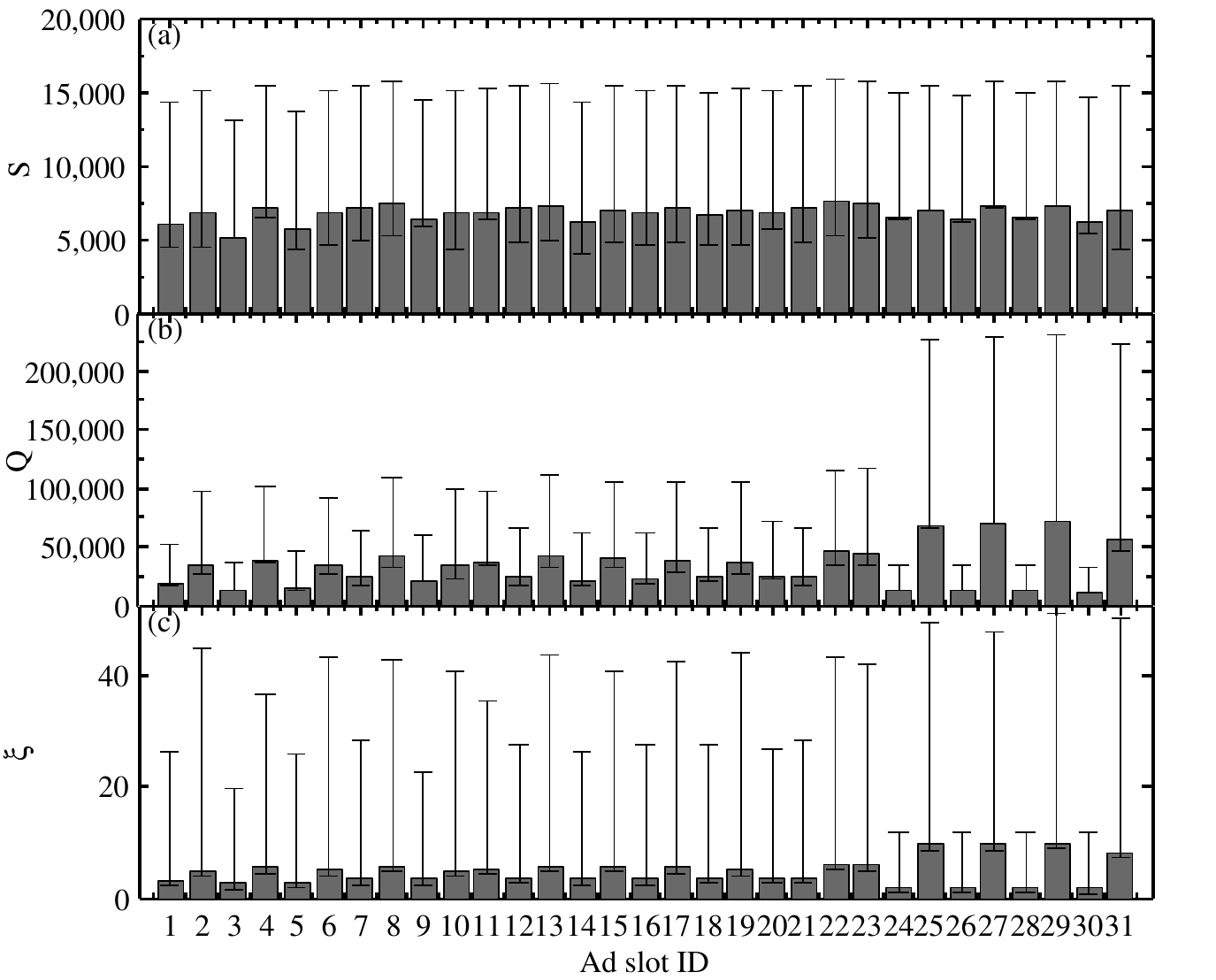}\vspace{-5pt}
\caption{The overview of daily supply and demand of ad slots in the SSP dataset in the training period: $S$ is the number of total supplied impressions; $Q$ is the number of total demand impressions; $\xi$ is the per impression demand (i.e., the number of advertisers who bid for an impression). }
\label{fig:supply_demand_ad_slots}
\vspace{7pt} \hspace{2pt}
\includegraphics[width=0.85\linewidth]{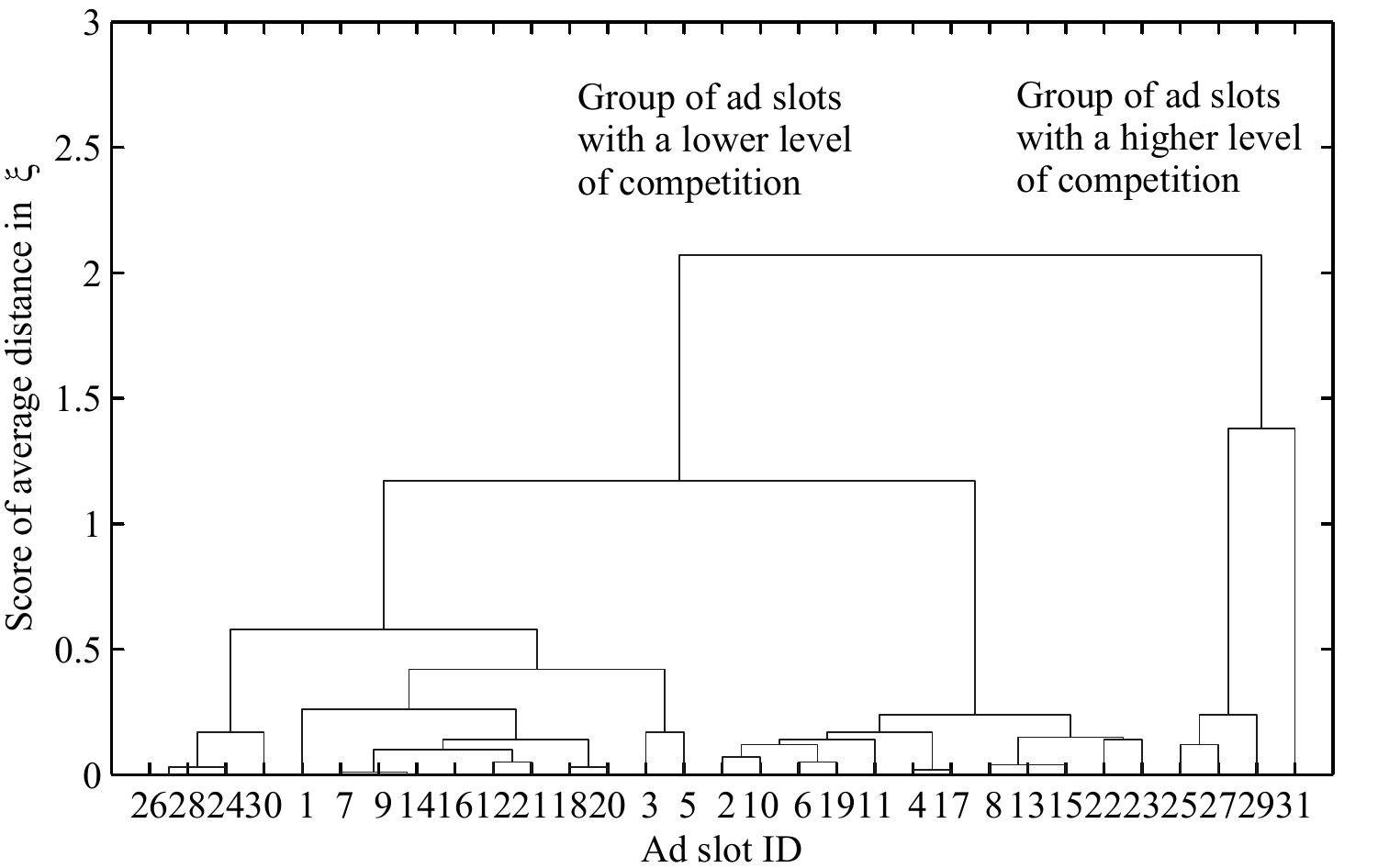}
\vspace{-7pt}
\caption{The hierarchical cluster tree of ad slots in the SSP dataset where the cluster metric is average distance in the per impression demand $\xi$.}
\label{fig:clustering}
\end{figure} 

Figure~\ref{fig:supply_demand_ad_slots} shows some descriptive statistics about supply and demand of all 31 ad slots from the SSP dataset in the training set. The ad slots have the same daily supply levels as well as their upper and lower bounds. However, the levels of daily demand are significantly different:~\texttt{AdSlot25}, \texttt{AdSlot27}, \texttt{AdSlot29} and \texttt{AdSlot31} are in higher demand than others, about 9 bidders per impression while the rest ad slots have the average value around 5. As shown in Figure~\ref{fig:clustering}, we take the average distance~\cite{Han_2011} in $\xi$ as the metric to cluster ad slots and obtain two groups. Note that $\xi$ significantly deviates from its mean value in a day's period because many more advertisers join RTB at peak time from 6am to 10am. In these hours, $\xi$ is 118.96\% higher than other hours. We can develop regression or time-series models to estimate $Q$ and $S$ on the delivery date; however, this is not a significant part of our study so we consider them as given parameters. 



\subsection{Bids and Payment Prices}\label{dp:exp_bid_and_payment_price}

\begin{table}[tp]
\scriptsize
\centering
\caption{Summary of bids distribution tests, where the numbers in the Kolmogorov-Smirnov (K-S) and Jarque-Bera (J-B) tests represent the percentage of tested auctions that have lognormal bids.}
\label{tab:bids_distribution_test}
\begin{tabular}{r|rrr}
\hline
  Group of ad slots  & \# of auctions  &  K-S test  &  J-B test \\
  & where $\xi \geq 30$ & & \\
\hline                  
Low competition  & 286   &  0.00\%  & 0.00\% \\
High competition & 15702 &  0.00\%  & 0.00\% \\
\hline
\end{tabular}
\vspace{2pt}
\caption{Comparison of estimations between the empirical distribution and the actual bids: $\phi(\xi)$ is the expected payment price; $\psi(\xi)$ is the standard deviation of payment prices; $\pi(\xi)$ is the expected winning bid; and the per impression demand $\xi = Q/S$.}
\label{tab:forecasting_by_sampling}
\begin{tabular}{r|rrr}
\hline
  Group of ad slots  & Difference     &  Difference     &  Difference     \\
                     & in $\phi(\xi)$ &  in $\psi(\xi)$ &  in $\pi(\xi)$ \\
\hline                  
Low competition      & 14.35\%  &  814.45\%  & 24.43\% \\
High competition     & 6.23\%   &   11.25\%  &  1.22\% \\ 
\hline
\end{tabular}
\vspace{5pt}
\end{table}


Once $\xi$ is given, we can use either probabilistic or empirical model to estimate the corresponding payment price $\phi(\xi)$ in RTB. For the probabilistic model, we consider symmetric bidders and assume their bids follow a log-normal distribution. However, the distribution tests shown in Table~\ref{tab:bids_distribution_test} reveal the fact that actual bids in RTB are not log-normally distributed. This confirms the statement that advertisers in the real-world are not symmetric. They may frequently change their bids for unclear reasons. Therefore, we use the empirical method to estimate $\phi(\xi)$ as well as $\psi(\xi)$ and $\pi(\xi)$.

\begin{figure}[t]
\scriptsize
\centering
\includegraphics[width=0.975\linewidth]{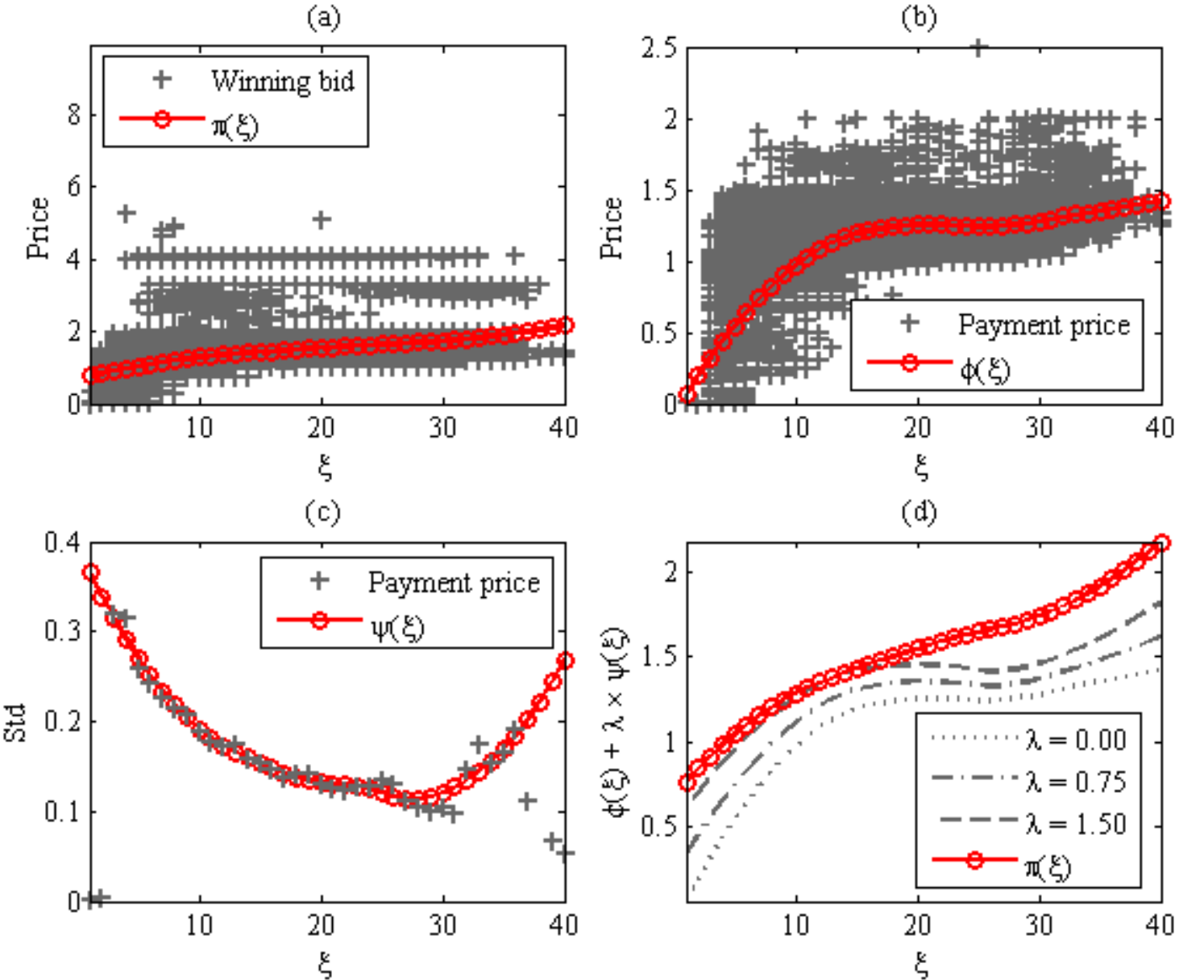}
\vspace{-7pt}
\caption{An empirical example of estimating $\pi(\xi)$, $\phi(\xi)$ and $\psi(\xi)$ for \texttt{AdSlot25} from historical bids, where $\xi$ is the per impression demand, $\pi(\xi)$ is the expected winning bid, $\phi(\xi)$ is the expected payment price and $\psi(\xi)$ is the standard deviation of payment prices.}
\label{fig:advertiser_cost}
\end{figure}

Figure~\ref{fig:advertiser_cost} illustrates an example of our empirical distribution method for \texttt{AdSlot25}. In the learning set, each winning price can be plot against the demand level $\xi$. We then use Algorithm~\ref{algo:rlwr} to compute $\phi(\xi)$. As described earlier, $\psi(\xi)$ and $\pi(\xi)$ are obtained numerically in the similar manner. In our experiments, we allow 10\% span of smoothing. As shown in Figure~\ref{fig:advertiser_cost}, $\phi(\xi)$ and $\pi(\xi)$ are increasing with $\xi$ while $\psi(\xi)$ shows a quadratic pattern on $\xi$. Once $\xi$ is given, we can calculate the value of the terminal condition $p(0)$ by Eq.~(\ref{eq:constraint_1}). Figure~\ref{fig:advertiser_cost} also confirms our earlier statement on $\lambda$. Advertisers are not risk-averse if $\lambda = 0$; and they are risk-sensitive for a large $\lambda$. In our experiments, we set $\lambda = 1$. 

Table~\ref{tab:forecasting_by_sampling} examines the forecast performance of empirical method and compares the estimated values of $\phi(\xi)$, $\psi(\xi)$, $\pi(\xi)$ to the results of actual bids in the test set. The estimations of $\phi(\xi)$ and $\pi(\xi)$ are much better accurate than that of $\psi(\xi)$. We find that the weak estimations of  $\psi(\xi)$ mainly come from \texttt{AdSlot24}, \texttt{AdSlot26}, \texttt{AdSlot28} and \texttt{AdSlot30}. Their average per impression demand (in both learning and test sets) are around 1.3. As also shown in Figure~\ref{fig:advertiser_cost}, the lower $\xi$ the larger $\psi(\xi)$. Therefore, for the ad slots with a very low level of competition, we set $p(0)  = \pi(\xi)$. 



\subsection{Demand for Guaranteed Impressions}\label{dp:exp_demand_guaranteed_imps}

The advertisers' purchase behavior of guaranteed impressions is modeled by parameters $\alpha$, $\beta$, $\zeta$, $\eta$ as well as be restricted by the expected risk-aversion cost $\phi(\xi) + \lambda \psi(\xi)$. Here we discuss how to learn the values of $\alpha$ and $\zeta$. 

\begin{figure*}[t]
\centering
\begin{minipage}{0.475\linewidth}
\scriptsize
\centering
\includegraphics[width=0.9\linewidth]
{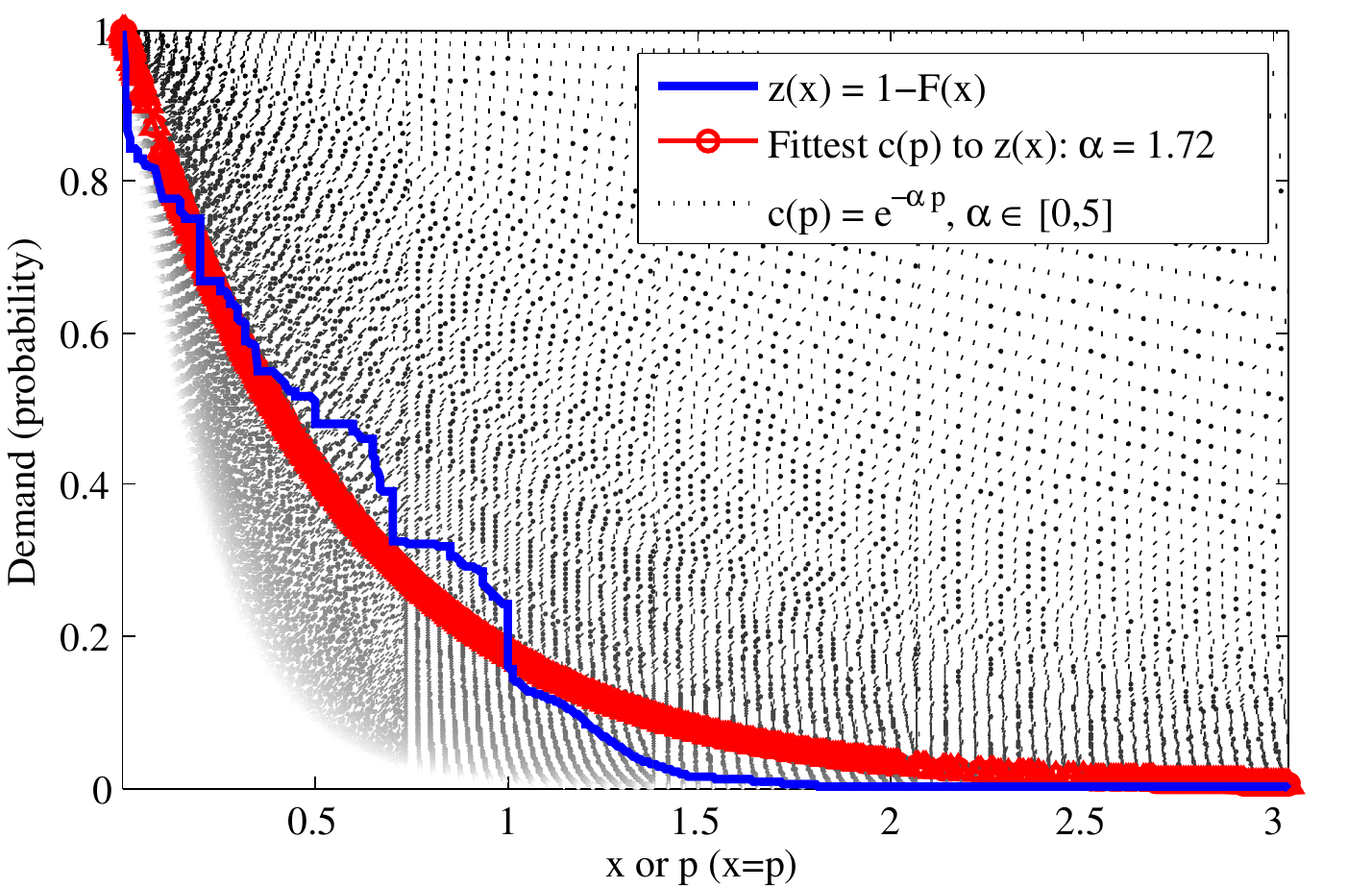}
\vspace{-5pt}
\caption{An empirical example of estimating the value of $\alpha$ for \texttt{AdSlot25}, where $\alpha$ is calculated based on the smallest RMSE between the inverse function of empirical CDF of bids $z(x) = 1- F(x)$ and the function $c(p)=e^{-\alpha p}$.}
\label{fig:demand_price_calibration} 
\end{minipage}
\begin{minipage}{0.475\linewidth}
\scriptsize
\centering
\includegraphics[width=0.9\linewidth]{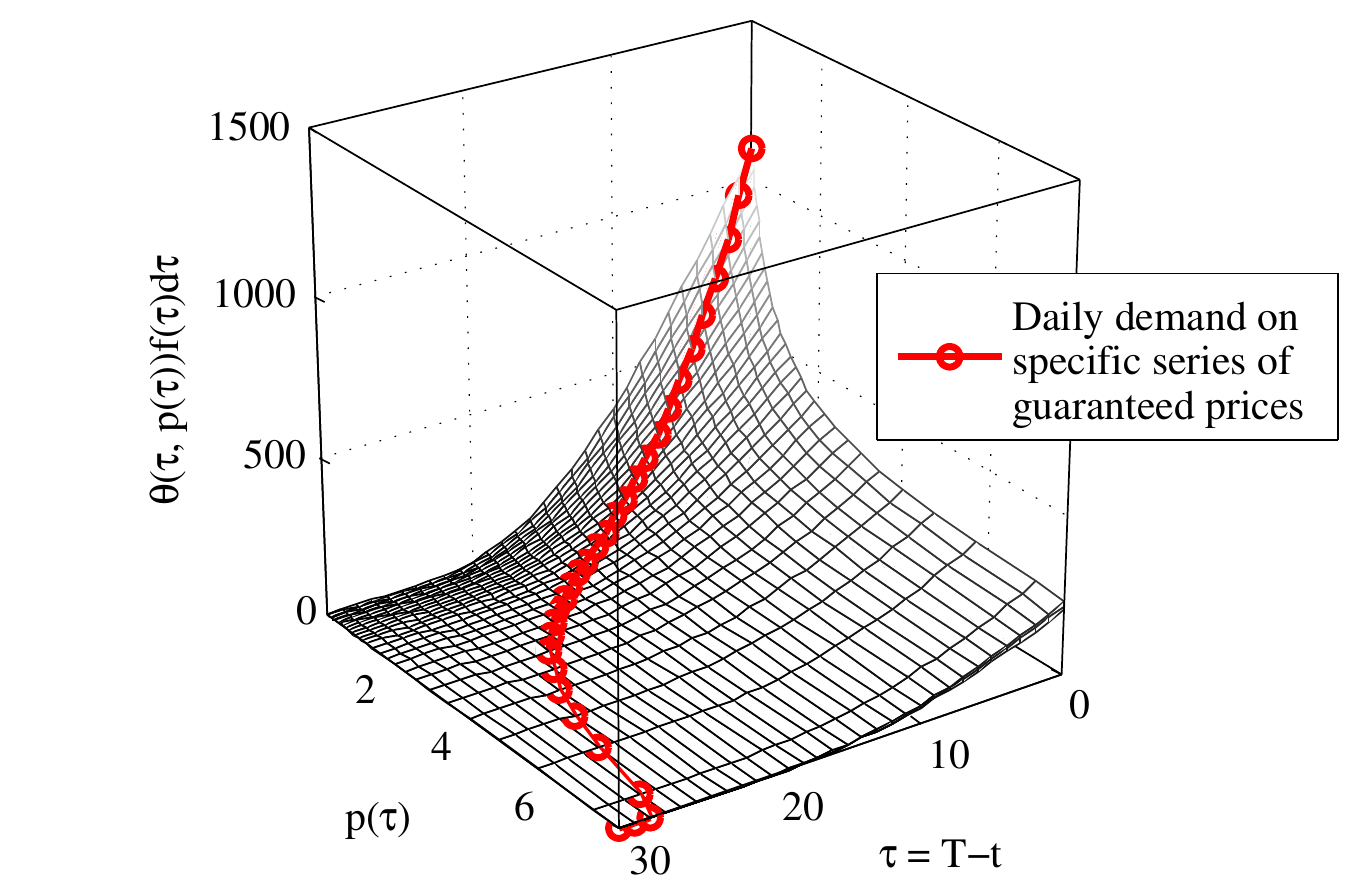}
\vspace{-7pt}
\caption{A numerical example of demand surface for guaranteed impressions of an ad slot, where $\theta(\uptau, p(\uptau)) f(\uptau) d \uptau$ represents the number of advertisers who will buy guaranteed impressions at $p(\uptau)$ and in $[\uptau, \uptau + d\uptau]$; other parameters are $\alpha=1.85; \beta=0.01; \zeta=2000; \eta=0.01; T=30$.}
\label{fig:demand_anlaysis}
\vspace{-5pt}
\end{minipage}
\begin{minipage}{0.85\linewidth}
\begin{table}[H]
\scriptsize
\centering
\caption{Summary of plot notations in Figures~\ref{fig:reve_analysis_1}~\&~\ref{fig:reve_analysis_2}.}
\label{tab:reve_notations}
\begin{tabular}{rp{5in}}
\hline
\multicolumn{2}{l}{Calculated revenue:}\\
\texttt{R-I} & The optimal total revenue calculated based on the estimated demand; \\
\texttt{R-II} & The optimal total revenue calculated based on the actual bids in the test set; \\
Baseline:\\
\texttt{B-I}    & The RTB revenue calculated based on the actual winning bids (1st price) in the test set; \\
\texttt{B-II}   & The RTB revenue calculated based on the actual payment (2nd) price in the test set; \\
\texttt{B-III}  & The estimated RTB revenue based on the learned empirical distribution; \\
\hline
\end{tabular}
\caption{Summary of revenue evaluation of all 31 ad slots in the SSP dataset.}
\label{tab:reve_summary}
\begin{tabular}{c|rrr|rr}
\hline
\multirow{5}{*}{Group of ad slots}      
& \multicolumn{3}{c|}{Performance of revenue maximization}      
&  \multicolumn{2}{c}{Performance of price discrimination} \\
\cline{2-6} 
         & Estimated     & Actual        & Difference of RTB             & Ratio of actual    &  Ratio of actual  \\
         & revenue        & revenue       & revenue between               & 2nd price reve     &  optimal reve   \\
		 & increase       & increase      & estimation \&                 & to actual          &  to actual  \\
		 &                &               & actual payment                & 1st price reve     &  1st price reve \\
\hline
Low competition   & 31.06\%  &  8.69\%   & 13.87\%  & 67.05\% & 81.78\% \\
High competition  & 31.73\%  & 21.51\%   & 6.23\%   & 78.04\% & 94.70\%\\
\hline
\end{tabular}
\vspace{5pt}
\end{table}
\end{minipage}
\end{figure*}

If we only consider the price effect, we can create the function $c(p) = e^{-\alpha p}$ from Eq.~(\ref{eq:demand_function_1}) to represent the probability that an advertiser would like to buy an impression at price $p$ when $\uptau =0$. In RTB, this probability can be learned from the data by investigating the inverse cumulative distribution function (CDF) of all bids, denoted by $z(x) = 1 - F(x)$. For a same domain space of $p$ and $x$, we can have two series of probabilities $c(p)$ and $z(x)$. Therefore, $\alpha$ can be calibrated as the value that gives the smallest root mean square error (RMSE) between $c(p)$ and $z(x)$. Figure~\ref{fig:demand_price_calibration} illustrates an empirical example of this calibration graphically for \texttt{AdSlot25} where the estimated $\alpha = 1.72$. 

The values of $\zeta$ can also be calibrated from data. Consider a small time step $d \widetilde\uptau$, then we have the following inequality
\begin{align}
e^{-\alpha p ( 1 + \beta \times 0 )} \zeta e^{-\eta \times 0 } d \widetilde\uptau & \ \leq Q \times (1 - F(\{x >= p\})).
\end{align}
If $d \widetilde\uptau = 1$, then we can have $\zeta = Q \times (1 - F(\{x >= p\})) / e^{-\alpha p}$.

It is difficult to learn the values of parameters $\beta$ and $\eta$ given our current datasets. The two parameters represent the time effect on advertiser\rq{}s buy behavior of guaranteed impressions. Here we simply adopt the initial parameter settings used in the flight tickets booking system~\cite{Anjos_2004,Malighetti_2009} and set $\beta = \eta = 0.2$. These two parameters can be then updated if the PG system runs for a certain period of time. By having the values of all the model parameters, we can construct the demand surface for a certain range of price series. Figure~\ref{fig:demand_anlaysis} presents a demand surface that satisfies \textbf{A-1} and \textbf{A-2}. It is convex in the guaranteed price and in the time interval between the purchase time and the delivery date. 

\subsection{Revenue Analysis}\label{dp:exp_reve_analysis}

We first present two empirical examples to illustrate how the developed model works with different levels of competition and then provide the overview results. 

\begin{figure*}[t]
\begin{minipage}{0.485\linewidth}
\small
\centering
\includegraphics[width=0.95\linewidth]{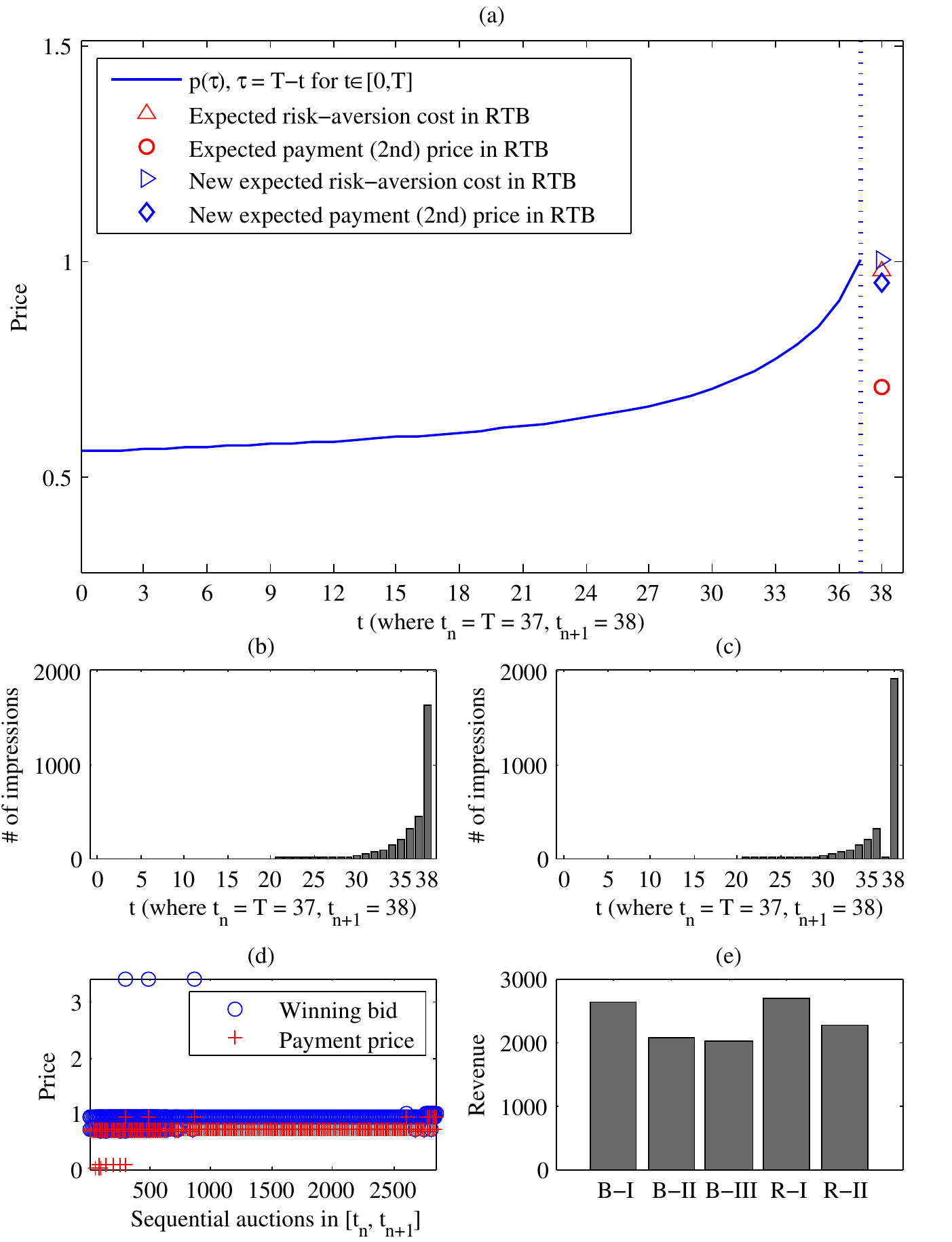}\vspace{-10pt}
\caption{An empirical example of \texttt{AdSlot14}: (a) the optimal dynamic guaranteed prices; (b) the estimated daily demand; (c) the daily demand calculated based on the actual bids in RTB on the delivery date; (d) the winning bids and payment prices in RTB on the delivery date; (e) the comparison of revenues [see Table~\ref{tab:reve_notations} for \texttt{B-I},  \texttt{B-II}, \texttt{B-III}, \texttt{R-I}, \texttt{R-II}]. The parameters are: $Q = 17691; S=2847; \alpha = 2.0506; \beta = 0.2; \zeta = 442; \eta = 0.2; \omega = 0.05; \kappa=1; \gamma = 0.4240$; $\lambda = 2$.}
\label{fig:reve_analysis_1}
\end{minipage}
\begin{minipage}{0.02\linewidth}
\end{minipage}
\begin{minipage}{0.485\linewidth}
\centering
\includegraphics[width=0.95\linewidth]{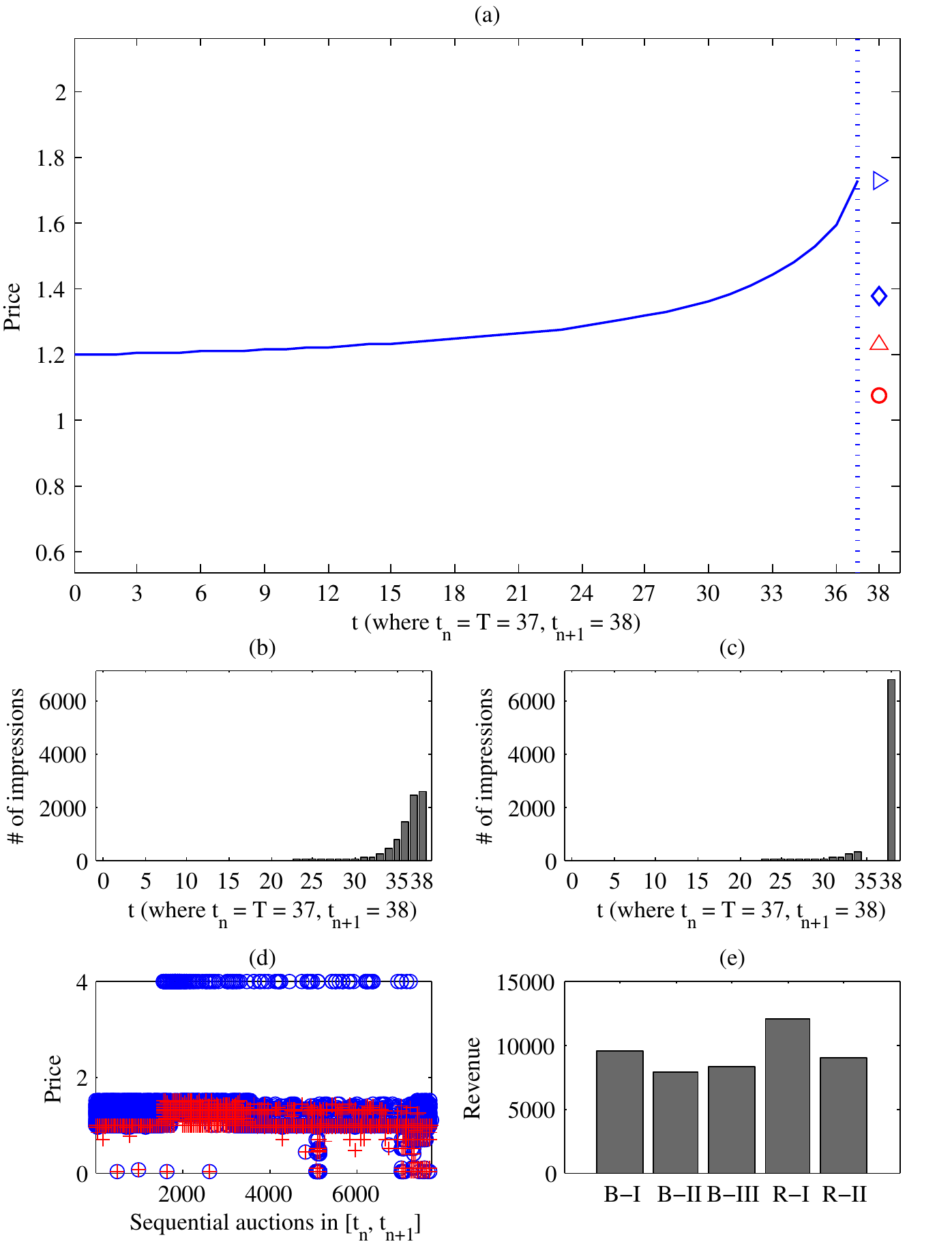}\vspace{-10pt}
\caption{An empirical example of \texttt{AdSlot27}: (a) the optimal dynamic guaranteed prices; (b) the estimated daily demand; (c) the daily demand calculated based on the actual bids in RTB on the delivery date; (d) the winning bids and payment prices in RTB on the delivery date; (e) the comparison of revenues [see Table~\ref{tab:reve_notations} for \texttt{B-I},  \texttt{B-II}, \texttt{B-III}, \texttt{R-I}, \texttt{R-II}]. The parameters are: $Q = 89126; S=7678; \alpha = 1.7932; \beta = 0.2; \zeta = 2466; \eta = 0.2; \omega = 0.05; \kappa=1; \gamma = 0.66$; $\lambda = 2$.}
\label{fig:reve_analysis_2}
\end{minipage}
\end{figure*}

Figure~\ref{fig:reve_analysis_1} shows an example of a less competitive market. The learned average per impression demand on \texttt{AdSlot14} is about $3.39$ (in the test set the actual $\xi = 6.21$). In such a market, advertisers would be less willing to purchase future impressions in advance because they think they can obtain the targeted impressions at lower payment prices. The model finally allocates 42.40\% of future impressions to the guaranteed contracts. In the meantime, the calculated guaranteed prices are not expensive. The prices start with a value lower than the expected payment price from RTB and steadily increases into the level that is close to the maximum value of advertisers' bids. In Figure~\ref{fig:reve_analysis_1}, we find that our forecasting values are close to the actual campaigns because the estimated RTB revenue \texttt{B-II} is almost as same as the actual RTB revenue \texttt{B-III}. Therefore, the estimated advertisers' demand for guaranteed impressions arrives are as similar as the actual daily demand (see Figure~\ref{fig:reve_analysis_1}(b)\&(c)). We also test the guaranteed selling with the actual bids in the test set and find that the calculated revenue \texttt{R-II} is still higher than actual second-price RTB revenue \texttt{B-II}. This shows that the developed model successfully segments advertisers.

Figure~\ref{fig:reve_analysis_2} describes an example of a competitive market, where the learned average per impression demand for \texttt{AdSlot27} is 9.63 (in the test set the actual $\xi = 11.61$). More advertisers would be willing to purchase guaranteed impressions in advance because of the increased level of competition and risk. The model finally allocates 66\% future impressions to guaranteed contracts and suggests higher prices at the beginning of guaranteed selling. The estimated total revenue is maximized (i.e., \texttt{R-I} $>$ \texttt{B-III}); the optimal revenue calculated by the actual bids is more than the actual second-price RTB revenue (i.e., \texttt{R-II} $>$ \texttt{B-II}). 

The overall results are presented in Table~\ref{tab:reve_summary}. The revenues calculated based on the estimated demand are always maximized. If we use the actual bids to calculate the demand at the given guaranteed prices, we can also have increased revenues (compared to the actual second-price auction market). The results successfully validate the developed model and we find the model performs better in a high competitive environment. This is because the publisher's revenue is actually maximized by the price discrimination, and in a competitive market there are more risk-averse advertisers to segment. With more and more risk-averse advertisers buying the guaranteed impressions, the increased total revenue will approximate advertisers' private evaluations.

\section{Concluding Remarks}\label{dp:conclusion}

We investigate a revenue maximization model for a publisher (or SSP) who engages in RTB to provide guaranteed delivery of display impressions. We not only design the mechanism tailored to RTB but also explore its feasibility and performance by mining the real datasets. Our experimental evaluation successfully validates the developed model as the publisher receives increased revenues. This work opens several directions for future research. First, we can further consider stochastic supply and demand. Second, a parametric updating framework for multi-period pricing and allocation would be of interest.

{
\balance
\scriptsize
\bibliographystyle{abbrv}
\bibliography{pg}
\balancecolumns
}
\end{document}